\documentclass[fleqn,usenatbib,useAMS,twocolumn,a4paper,aps]{revtex4}

\usepackage{aas_macros}
\usepackage[T1]{fontenc}
\usepackage{ae,aecompl}
\usepackage{hyperref}
\usepackage{soul}
\usepackage{xcolor}
\usepackage{ragged2e}

\usepackage{graphicx}	
\usepackage{amsmath}	
\usepackage{amssymb}	
\usepackage{bm}		
\usepackage{pdflscape}	

\begin{document}



\title[Massloss from magnetized disk]{Study of mass outflows from magnetized accretion disks around rotating black holes with thermal conduction}

\author{Camelia Jana$^1$}
\email{camelia\_jana@iitg.ac.in}

\author{Monu Singh$^1$}
\email{monu18@iitg.ac.in}

\author{Suvendu Rakshit$^2$}
\email{suvenduat@gmail.com}

\author{Santabrata Das$^1$}
\email{sbdas@iitg.ac.in}

\affiliation{$^1$Department of Physics, Indian Institute of Technology Guwahati, Guwahati, 781039, Assam, India.}
\affiliation{$^2$Aryabhatta Research Institute of Observational Sciences, Manora Peak, Nainital 263002, India}

\begin{abstract}

We examine mass outflows from a low-angular momentum, viscous, advective, and magnetized accretion disk around a rotating black hole in presence of thermal conduction. We consider the disk is primarily threaded by the toroidal component of the magnetic field and an effective potential satisfactorily mimicked the spacetime geometry around the rotating black hole. With this, we self-consistently solve the coupled governing equations for inflow and outflow and compute the mass outflow rate $R_{\dot m}$ (ratio of mass flux of inflow to outflow) in terms of the inflow parameters, namely energy ($\mathcal{E}$), angular momentum ($\lambda$), plasma-$\beta$ and conduction parameter ($\Upsilon_{\rm s}$) around weakly  rotating ($a_{\rm k} \rightarrow 0$) as well as rapidly rotating ($a_{\rm k} =0.99$) black holes. Our findings reveal that the present formalism admits coupled inflow-outflow solutions across a wide range of inflow parameters yielding substantial mass loss. We observe that $R_{\dot m}$ monotonically increases with $\Upsilon_{\rm s}$, irrespective of black hole spin. We also find that for a fixed $\Upsilon_{\rm s}$, when energy, angular momentum, and magnetic field strength of the inflowing matter is increased, $R_{\dot{\rm m}}$ is enhanced resulting the outflows even more pronounced. We further estimate the maximum outflow rate ($R^{\rm max}_{\dot{\rm m}}$) by varying the inflow parameters and find that thermal conduction leads to maximum mass outflow rate $R^{\rm max}_{\dot{\rm m}} \sim 25\%$ for rapidly rotating black hole of spin $a_{\rm k} = 0.99$. Finally, we employ our formalism to explain the kinetic jet power of $68$ radio-loud low-luminosity active galactic nuclei (LLAGNs), indicating that it is potentially promising to account for the observed jet power of substantial number of LLAGNs.

\end{abstract}

\maketitle
	

\section{Introduction}

The accumulation of matter onto black holes (BHs) powers various high-energy astrophysical phenomena in the universe, including X-ray binaries (XRBs), gamma-ray bursts (GRBs), and active galactic nuclei (AGNs). Several theoretical models have been developed to understand the nature of these accretion phenomena. The standard thin accretion disk model, introduced by \cite{Shakura-Sunyaev-1973}, assumes efficient radiative cooling, where  energy generated by viscous dissipation is immediately radiated away from the system, allowing the disk to remain cool. This yields a geometrically thin and optically thick disk. In contrast, the radiatively inefficient accretion flow (RIAF) model indicates that most of the energy generated is retained within the gas and is advected inward with the flow, rather than being radiated away.
The RIAF can be categorized on the basis of mass accretion rate and optical depth. At a very high accretion rate, the flow becomes optically thick, and photons are trapped within the gas, leading to a radiatively inefficient state described by the `slim accretion disk model' \cite[]{Abramowicz-etal-1988}. Conversely, at low accretion rate, the flow becomes optically thin and radiative cooling becomes inefficient, leading to an advection-dominated accretion flow (ADAF) \cite[]{Narayan-Yi-1994, Narayan-Yi-1995}. The ADAF model has successfully explained various observational features of low-luminosity active galactic nuclei (LLAGNs) and black hole X-ray binaries (BH-XRBs) \cite[]{Lasota-etal-1996,Manmoto-etal-1997, Esin-etal-1997, Hameury-etal-1997, Narayan-etal-1998, Menou-etal-1999,  Matteo-etal-2000, Melia-Falcke-2001, Yuan-Narayan-2014}.

Indeed, an intriguing aspect of advective accretion process is the presence of winds and outflows in LLAGNs and XRBs \cite[]{Wang-etal-2013,Homan-etal-2016,Park-etal-2019, Shi-etal-2022}.  These outflows seem to arise because of various physical mechanisms, including excess gas pressure, centrifugal forces, magnetic pressure, and radiation pressure \cite[]{Chelouche-Netzer-2005,Hawley-Krolik-2006,das-chattopadhyay-2008, Ohsuga-Mineshige-2011,Dihingia-etal2021,Aktar-etal2024}. Recent study \cite{Jana-Das-2024} has demonstrated that magnetized disks surrounding rapidly rotating black holes can efficiently eject matter from the disk in the form of outflows.

In low-density advective accretion flows, the electron mean free path is significantly larger than the size of the disk, rendering the plasma to remain collisionless \cite[]{Johnson-Quataert-2007, Tanaka-Menou-2006}. In this scenario, thermal conduction plays a vital role in transporting energy from the hot inner region to the cooler outer region, thereby significantly influencing the thermodynamic properties of the accretion flows. Meanwhile, several studies have investigated the effects of thermal conduction on accretion flows \cite[]{Tanaka-Menou-2006, Johnson-Quataert-2007, Shadmehri-2008, Faghei-2012, Khajenabi-Shadmehri-2013, Bu-etal-2016, Ghoreyshi-Shadmehri-2020, Mitra-etal-2023, Ghasemnezhad-Khosravi-2024, Singh-Das-2024b}. Notably, \cite{Tanaka-Menou-2006} proposed that thermal conduction can aid in launching outflows, while \cite{Bu-etal-2016} demonstrated that it enhances wind velocity including the increase of energy in manyfold. Furthermore, \cite{Khajenabi-Shadmehri-2013} found that thermal conduction affects the geometry of outflows by reducing their opening angle, suggesting that it helps in collimating the outflows. In a recent attempts, \cite{Rezgui-etal-2019, Rezgui-etal-2022} demonstrated that thermal conduction enhances jet efficiency with faster and collimated mass ejection. All these findings evidently indicate that thermal conduction seems to play a crucial role in ejecting matter from the accretion disk.

Accretion flows onto black holes exhibit complex dynamical behavior. Matter starting from large distances with negligible speed gains radial velocity, and eventually crosses the event horizon at the speed of light. Therefore, to meet the event horizon condition, any black hole solution must make smooth sonic state transition from subsonic to supersonic at the critical point. Depending on the initial conditions, the flow may have either a single or multiple critical points. The presence of multiple critical points is potentially necessary, as it can lead to shock transitions. 

As the rotating matter accretes inward, it experiences centrifugal repulsion, which slows down the flow and leads to the accumulation of matter around the black hole. At its threshold, this repulsion triggers a shock transition when the shock conditions are favorable \cite[]{Landau-Lifshitz-1959}. The shock causes a sudden compression of matter, forming a hot, dense post-shock corona (PSC) that acts as an effective boundary layer surrounding the black hole and emitting high-energy radiation. In the PSC, an excess thermal gradient force develops, deflecting a part of the infalling matter in the vertical direction as bipolar outflows or jets. This compelling mass loss mechanism from the accretion disk has also been validated by numerical studies \cite[]{Molteni-etal-1994, Molteni-etal1996, Das-etal-2014, Okuda-2014, Okuda-Das-2015, Okuda-etal2019}. Subsequently, several theoretical studies have explored various complexities of this process in recent years \cite[]{Chakrabarti-1999, Das-etal-2001a, chattopadhyay-Das-2007, das-chattopadhyay-2008, aktar-etal-2015, Kumar-Chattopadhyay2017, aktar-etal-2017, aktar-etal-2019, Jana-Das-2024}. However, the role of thermal conduction in generating outflows from the shocked accretion flow remains unexplored.

Motivated by this, we extend our previous work \cite{Jana-Das-2024}, and study the accretion-ejection mechanism in a low angular momentum, viscous, advective, magnetized accretion disk, incorporating the effects of thermal conduction. We consider accretion flow predominantly threaded by toroidal magnetic fields \cite[]{Oda-etal-2007} and adopt an effective potential to mimic the space-time around a rotating black holes \cite{Dihingia-etal2018a}. We self-consistently solve the coupled inflow and outflow equations around the rotating black hole of spin $a_{\rm k}$ and quantify the mass outflow rate ($R_{\dot{\rm m}}$) in terms of inflow parameters, namely conduction parameter ($\Upsilon_{\rm s}$), energy ($\mathcal{E}$), angular momentum ($\lambda$), and plasma-$\beta$ (measure of magnetic fields strength). Our findings indicate that thermal conduction significantly influences $R_{\dot{\rm m}}$; as thermal conduction increases, so does the outflow rate. Furthermore, we calculate the maximum outflow rate ($R^{\rm max}_{\dot{\rm m}}$) by freely varying the model parameters and observe that $R^{\rm max}_{\dot{\rm m}}$ remains higher for a rapidly rotating black hole ($a_{\rm k} = 0.99$) compared to a non-rotating black hole ($a_{\rm k} = 0.0$). Finally, we discuss the implications of our findings in the context of jet kinetic power observed in low luminosity AGNs (LLAGNs).

The paper is structured as follows. In Section \ref{section2}, we describe the model assumptions and governing equations. We present the obtained results in Section \ref{section3}. In Section \ref{section4}, we discuss the observational implication of the present formalism for LLAGNs. Finally, we summarize our findings in Section \ref{section5}.

\section{Assumptions and Governing equations}
\label{section2}

We consider an axisymmetric disk-jet system around a rotating BH in the steady state. Specifically, we assume that the disk is confined around the equatorial plane where the accretion occurs, and the jet geometry aligns with the BH rotational axis. Here, we use cylindrical coordinate system ($x$, $\phi$, $z$) with the BH located at origin, and $z = 0$ defines the equatorial plane. Further, to express all the governing equations, we choose the geometrical unit system where the mass of the black hole ($M_{\rm BH}$), universal gravitational constant ($G$), and speed of light ($c$) are chosen as unity ($M_{\rm BH} = G = c =1 $). In this system, length, angular momentum, and specific energy are expressed in the units of $GM_{\rm BH}/c^2$, $GM_{\rm BH}/c$, and $c^2$, respectively.

\subsection{Governing equations for accretion}

We begin with a low angular momentum, viscous, magnetized, advective accretion flow around a rotating BH, incorporating thermal conduction as the heat transfer mechanism. To model the magnetic field structure, we refer to numerical simulations \cite[]{Hirose-etal-2006, Machida-etal-2006, Johansen-Levin-2008}, which demonstrate that the magnetic fields inside the disk are turbulent in nature and primarily dominated by the toroidal component. Following these simulation works, the total magnetic fields can be decomposed into mean field, $B$ = ($0$, $\langle B_{\phi}\rangle$, $0$) and fluctuating field, $\delta B$ = ($\delta B_{x}$, $\delta B_{\phi}$, $\delta B_{z}$). Here, `$\langle$ $\rangle$' corresponds to the azimuthal average. Upon azimuthal averaging, we assume that the fluctuating terms vanish ($\langle \delta B \rangle = 0$). Therefore, the azimuthal component becomes dominant over the radial and vertical component, i.e, $\vert \langle B_{\phi} \rangle + \delta B_{\phi} \vert$ $>>$ $\vert \delta B_{x} \vert$ and $\vert \delta B_{z} \vert$. This ultimately leads to the azimuthally averaged magnetic field as $\langle B \rangle$ = $\hat{\phi}$ $\langle B_{\phi} \rangle$ \cite[]{Oda-etal-2007}. Further, to account for general relativistic effect, we adopt the recently introduced pseudo-potential \cite[]{Dihingia-etal2018a}, which satisfactorily replicates the space-time geometry around a rotating BH. The effective potential at the disk equatorial plane is given by,
\begin{equation}
     \Psi_{\textrm{e}} ^{\textrm{eff}} = \frac{1}{2}\ln\left[\frac{x \Delta}{x^3 + (a_{\rm k} + \lambda)(a_{\rm k} - \lambda)x+ 2 (a_{\rm k} + \lambda)^2} \right],
     \label{potequi}
\end{equation}
where $\lambda$ is the local specific angular momentum of inflow, $a_{\rm k}$ is the spin parameter of BH, and $\Delta = x^2 - 2 x + a_{\textrm{k}}^2$. Based on the above considerations, the governing magnetohydrodynamic (MHD) equations \cite[]{Sarkar-Das2016, Sarkar-etal2018, Dihingia-etal-2020} describing the motion of the accretion flow are as follows:\\

\noindent (a) The radial momentum equation:
\begin{equation}
	v \frac{dv}{dx}+\frac{1}{ \rho}\frac{dP}{dx}+\frac{d \Psi_{\rm{e}}^{\rm{eff}}}{dx} + \frac{\langle B_{\phi}^2 \rangle}{4 \pi x \rho}  = 0,
 \label{ineng}
\end{equation}
Here, $x$, $v$, and $\rho$ represent the radial distance, radial velocity, and mass density, respectively. The total isotropic pressure $P$ combines both gas pressure ($P_{\rm gas}$) and magnetic pressure ($P_{\rm mag}$), such that $P = P_{\rm gas} + P_{\rm mag}$. For simplicity, we consider that ions are predominantly protons, and the electron number density ($n_{\rm e}$) equals the ion number density ($n_{\rm p}$), leading to $n_{\rm e} = n_{\rm p}$. Subsequently, assuming a single temperature for both ions and electrons, we express $P_{\rm gas} = 2 \rho \Theta/\tau$, where $\tau = 1 + m_{\rm p}/m_{\rm e}$ and $\Theta$ is the dimensionless temperature given as $\Theta = k_{\rm B} T/{m_{\rm e}c^2}$. Here, $k_{\rm B}$ is the Boltzmann constant, $T$ is the temperature in Kelvin, and $m_{\rm e}$ and $m_{\rm p}$ are the electron and proton masses, respectively. Further, the magnetic pressure is obtained as $P_{\rm mag} = \frac{\langle B_{\phi}^2 \rangle}{8 \pi}$, where $\langle B_{\phi}^2 \rangle$ denotes the azimuthal average of the square of toroidal component of the magnetic fields. We define the plasma-$\beta$ as the ratio of gas pressure to magnetic pressure, given by \(\beta = P_{\rm gas}/P_{\rm mag}\), and estimate the magnetic strength as $\langle B_{\phi}^2 \rangle = 8 \pi P_{\rm gas}/ \beta$. \\

\noindent (b) The mass conservation equation:
\begin{equation}
	\dot{M} = 2\pi v \Sigma \sqrt{\Delta}\thickspace,
 \label{inmass}
\end{equation} 
Here, $\dot{M}$ denotes the mass accretion rate, which remains constant everywhere except the region of mass loss, and $\Sigma$ refers to the vertically averaged surface mass density of the accreting matter \cite[]{Matsumoto-etal-1984}. Here, $\Sigma = 2 \rho H$, where $H$ denotes the local half thickness of the disk. Following \cite{Riffert-Herold-1995, peitz-Appl-1997}, we estimate $H$ as,
\begin{equation}
    H = \sqrt{\frac{Px^3}{\rho \mathcal{F}}},\quad \mathcal{F} =\frac{1}{(1-\lambda \Omega)} \times \frac{(r^2 + a_{\textrm{k}}^{2})^{2} + 2 \Delta a_{\textrm{k}}^{2}}{(x^2 + a_{\textrm{k}}^{2})^{2} - 2 \Delta a_{\textrm{k}}^{2}},
\end{equation}
where $\Omega$ being the angular velocity of the flow and is given by, $\Omega = ({2 a_{\textrm{k}} + \lambda (x - 2)})/({a_{\textrm{k}}^{2}(x + 2) - 2 a_{\textrm{k}}\lambda + x^3})$.\\

\noindent(c) The azimuthal momentum equation:
\begin{equation}
	v \frac{d\lambda}{dx} + \frac{1}{\Sigma\space x}\frac{d}{dx}(x^2 T_{x\phi}) = 0,
 \label{inlam}
\end{equation}
In equation (\ref{inlam}), we consider that the vertically integrated total stress is dominated by the ${x\phi}$-component of the Maxwell stress ($T_{x\phi}$) over the other components. Following \cite{Machida-etal-2006}, we calculate $T_{x\phi}$ in presence of significant radial motion of the accreting matter as \cite[]{chakrabarti-1996},
\begin{equation}
    T_{x\phi}= \frac{\langle B_{x} B_{\phi} \rangle}{4 \pi} H =  - \alpha_{\rm{B}} (W + \Sigma v^2),
\end{equation}
where $W ~(= 2 P H)$ is the vertically integrated total pressure \cite[]{Matsumoto-etal-1984}, and $\alpha_{\rm B}$ (ratio of Maxwell stress to the total pressure) is the constant of proportionality commonly known as viscosity parameter.\\

\noindent (d) The entropy generation equation:
\begin{equation}
	\frac{v \Sigma}{\Gamma - 1}\left(\frac{1}{\rho}\frac{dP_{\rm gas}}{dx} - \frac{\Gamma P_{\rm gas}}{\rho^2}\frac{d\rho}{dx}\right) = Q^{-}-Q^{+} - Q^{\rm cond},
 \label{entro}
\end{equation}
where $\Gamma$ is the adiabatic index, and $Q^{+}$ and $Q^{-}$ are the vertically integrated heating and cooling rates, respectively. Meanwhile, numerical simulation studies suggest that the flow is heated because of the thermalization of magnetic energy through the magnetic reconnection process \cite[]{Machida-etal-2006, Hirose-etal-2006}. Following this, the heating rate is calculated as,
\begin{equation}
	Q^{\rm +}= \frac{\langle B_{x} B_{\phi} \rangle}{4 \pi} x H \frac{d \Omega}{d x}= -\alpha_{\rm{B}}   \left(W + \Sigma v^2\right) x \frac{d \Omega}{d x}.
\end{equation} 
In equation (\ref{entro}), $Q^{\rm cond}$ accounts for the energy transfer resulting from saturated thermal conduction in collisionless plasmas. In reality, at low accretion rates, the electron mean free path becomes significantly larger than the length scale of the accretion flow, allowing the plasma to be treated as collisionless \cite[]{Tanaka-Menou-2006}. Based on this, $Q^{\rm cond}$ is given by  \cite[]{Cowie-Mckee-1977},
\begin{equation}
	Q^{\rm{cond}} = - \frac{2 H}{x} \frac{d (x F_{s})}{d x},
\end{equation}
where $F_{\rm s}$ represents the saturated conduction flux, defined as $F_{\rm s} = 5 \Upsilon_{\rm{s}} \rho \left(\frac{P_{\rm gas}}{\rho}\right)^{3/2}$. Here, $\Upsilon_{\rm{s}}$ is the dimensionless saturated conduction parameter (hereafter conduction parameter) that regulates the effect of thermal conduction within the disk and it takes values in the range $ 0 \leq \Upsilon_{\rm s} < 1$. Furthermore, since this study focuses on radiatively inefficient hot accretion flows characterized by low accretion rates, we neglect the cooling effects and  set $Q^{-}= 0$.\\

\noindent (e) Radial advection of the toroidal magnetic flux:
\begin{equation}
	\frac{\partial \langle B_{\phi} \rangle \hat{\phi}}{\partial t} = \nabla \times \left( \vec{v} \times \langle B_{\phi} \rangle \hat{\phi} - \frac{4\pi}{c} \eta \vec{j} \right).
 \label{flux}
\end{equation}
Here, $\vec{v}$ is the velocity vector, $\eta$ denotes the resistivity and the current density $\vec{j}$ is given by $\vec{j} = c(\nabla \times \langle B_{\phi}\rangle \hat{\phi})/4\pi$. Generally, the Reynold number for accretion disk is very high due to its large extent, and therefore, we neglect the magnetic diffusion term. Further, we ignore the dynamo term in this work. Finally, after taking the vertical average of the obtained equation with the condition that the azimuthally averaged magnetic field vanishes at the disk surface, we obtain the advection rate of the toroidal magnetic flux as \cite{Oda-etal-2007},
\begin{equation}
\dot{\Phi}_{\rm B} = - \sqrt{4\pi}v H {B} (x, z =0),
\label{mgflux}
\end{equation}
where $B(x, z=0)$ represents the azimuthally averaged toroidal magnetic field confined on the disk equatorial plane. In a realistic scenario, $\dot{\Phi}_{\rm B}$ does not remain constant but rather varies with $x$ due to the influence of the dynamo and magnetic diffusion terms. However, incorporating these effects is highly complex and falls outside the scope of this paper. Hence, following the work of \cite{Machida-etal-2006} in the quasi-steady state, we have, $\dot{\Phi}_{\rm B} \propto 1/x$. This approximation eventually captures the basic governing behavior of the advection of magnetic flux while avoiding the intricacies introduced by additional physical processes.

We close the governing equations (\ref{ineng} - \ref{flux}) of the accretion flow using an equation of state (EoS). In reality, the flow is thermally non-relativistic ($\Gamma = 5/3$) at the outer regions, while it becomes thermally relativistic ($\Gamma = 4/3$) at the vicinity of the black hole \cite[]{Frank-etal-2002}. Hence, a constant $\Gamma$ would not accurately describe the thermal variations of the accretion flow. Instead, it is more appropriate to compute $\Gamma$ self-consistently based on the disk temperature. Keeping this in mind, we adopt a relativistic equation of state (REoS), which is given by \cite[]{Chattopadhyay-Ryu-2009},
\begin{equation}
	\epsilon  = \frac{\rho}{\tau} f, 
 \label{eos}
\end{equation}
where $\epsilon$ denotes the internal energy of the flow and $f$ is defined in terms of the temperature as,
\begin{equation}
	f = \left[1 + \Theta \left(\frac{9\Theta + 3}{3\Theta+2}\right) \right] + \left[\frac{m_{\rm p}}{m_{\rm e}} + \Theta \left(\frac{9\Theta m_{\rm e} + 3 m_{\rm p}}{3\Theta m_{\rm e} + 2 m_{\rm p}}\right) \right].
    \label{req2}
\end{equation}
With this REoS, we calculate the polytropic index $N = \frac{1}{2}\frac{df}{d\Theta}$, adiabatic index $\Gamma = 1 + \frac{1}{N}$, and sound speed $C_{\rm s}^{2} = \Gamma P_{\rm gas}/(\epsilon + P_{\rm gas})$ \cite[]{Dihingia-etal-2020}. 

Using equations (\ref{ineng} - \ref{entro}), (\ref{mgflux}) and (\ref{eos}) and after some simple algebra, we obtain the wind equation which is given by,
\begin{equation}
    \frac{dv}{dx} = \frac{\mathcal{N} (x, v, \Theta, \lambda, \beta)}{\mathcal{D} (x, v, \Theta, \lambda, \beta)},
    \label{dvdx}
\end{equation}
where numerator ($\mathcal{N}$) and denominator ($\mathcal{D}$) are the explicit functions of the flow variables, which are detailed in Appendix A. Utilizing equation (\ref{dvdx}), we obtain the derivatives of $\Theta$, $\lambda$ and $\beta$ as,
\begin{equation}
    \frac{d\Theta}{dx} = \Theta_{0} + \Theta_{\rm v} \frac{dv}{dx},
    \label{dldx}
\end{equation}
\begin{equation}
    \frac{d\lambda}{dx} = \lambda_{0} + \lambda_{\rm v} \frac{dv}{dx},
    \label{dldx}
\end{equation}

\begin{equation}
    \frac{d\beta}{dx} = \beta_{0} + \beta_{\rm v} \frac{dv}{dx},
    \label{dbdx}
\end{equation}
where the coefficients $\Theta_{0}$, $\Theta_{\rm v}$, $\lambda_{0}$, $\lambda_{\rm v}$, $\beta_{0}$, and $\beta_{\rm v}$ are the explicit function of flow variables and given in Appendix A.

In the accretion process, a subsonic flow at the outer edge ($x_{\rm edge}$) of the disk gradually picks up speed and ultimately crosses the event horizon at light speed before plunging into the black hole. To meet the inner boundary condition, an accretion solution around a black hole must pass through at least one critical point ($x_{\rm c}$), where the sonic transition takes place. At $x_{\rm c}$, both $\mathcal{N}$ and $\mathcal{D}$ vanish simultaneously ($i.e.$, $\mathcal{N} = 0$ and $\mathcal{D} = 0$) and hence the gradient of radial velocity takes the form $\frac{dv}{dx}\vert_{x_{\rm c}} = \frac{0}{0}$. Since the flow remains smooth everywhere during its journey, the value of $\frac{dv}{dx}$ must be real and finite all throughout. Therefore, we apply l'H$\hat{\rm o}$pital's rule at $x_{\rm c}$ to calculate $(dv/dx)_{x_{\rm c}} = (\frac{d\mathcal{N}/dx}{d\mathcal{D}/dx})\vert_{ x_{\rm c}}$. In general, we get two values of  $(dv/dx)$ at ${x_{\rm c}}$. For a physically acceptable accretion solution, the flow necessarily passes through a saddle-type critical point, where both values of $(dv/dx)_{x_{\rm c}}$ are real and of opposite sign \cite[]{Kato-etal-1993,Das-2007}. Interestingly, depending on the inflow parameters, a flow can possess multiple critical points. The critical point closest to the black hole is referred to as the inner critical point ($x_{\rm in}$), while the furthest one is called the outer critical point ($x_{\rm out}$) \cite[]{Chakrabarti1990,Chakrabarti-Das2004,Das-2007,Dihingia-etal2019a}. In this work, we we consider $\alpha_{\rm B}$, $\Upsilon_{\rm s}$, and $a_{\rm k}$ as global parameters, while boundary values of $\lambda$ and $\beta$ at $x_{\rm c}$ are treated as local parameters. Using these input parameters, we perform the critical point analysis at $x_{\rm c}$ and determine the velocity $v_{\rm c}$ and temperature $\Theta_{\rm c}$ of the inflowing matter. Utilizing $v_{\rm c}$ and $\Theta_{\rm c}$, we integrate equations (\ref{dvdx}-\ref{dbdx}) from $x_{\rm c}$ inward up to the event horizon ($x_{\rm h}$) and then outward up to the outer edge ($x_{\rm edge}$) of the disk. Finally, we join these two parts to  obtain the complete global transonic accretion solution. Needless to say, the global solution allows us to determine the inflow variables at the outer edge of the disk at the $x_{\rm edge}$, namely $\mathcal{E}_{\rm edge}, \lambda_{\rm edge}$, and $\beta_{\rm edge}$ \cite[references therein]{Jana-Das-2024}. It is essential to emphasize that these inflow variables at $x_{\rm edge}$ yield identical global transonic solutions when one integrates equations (\ref{dvdx}-\ref{dbdx}) from the outer edge.  Furthermore, utilizing the local inflow variables, we approximate the local flow energy in the weak field limit as $\mathcal{E} \sim v^2/2 + C^2_{\rm s}/(\Gamma-1) + \left < B^2_{\phi} \right >/(4\pi \rho) + \Psi_{\rm e}^{\rm eff}$ .

\subsection{Governing equations for outflows}

The complex interaction between accretion and ejection near black holes provides insights into mass loss activities, and hence we assume that outflows originate from the accretion disk and move along the black hole rotation axis. As a part of the inflowing matter deflects as outflows, leading to mass loss from the system, these outflows remain tenuous in nature. Because of this, we disregard the differential rotation and neglect viscosity in the outflows. In addition, as the toroidal component of the magnetic field is assumed to be dominant and the matter is being ejected vertically, we neglect the magnetic fields within the outflows for simplicity. Furthermore, we adopt the polytropic equation of state, $P_{\rm j} = K_{\rm j} \rho^{\Gamma}_{\rm j}$, where the subscript `$\rm j$' represents jet variables, $K_{\rm j}$ measures the entropy of the outflowing matter and $\Gamma$ is the adiabatic index which is determined based on temperature using the REoS (equation \ref{eos}, \ref{req2}). Based on these considerations, the governing equations for outflows are as follows:

\noindent (I) The energy conservation:
\begin{equation}
    \mathcal{E}_{\rm j} = \frac{1}{2}v^2_{\rm j}  +  C^2_{\rm sj}/(\Gamma -1) + \Psi^{\rm eff},
    \label{outeng}
\end{equation}
and (II) the mass conservation:
\begin{equation}
    \dot{M}_{\rm out} = \rho_{\rm j} v_{\rm j} \mathcal{A}_{\rm j}.
    \label{outmass}
\end{equation} 
In equation (\ref{outeng}), $\mathcal{E}_{\rm j}$, $v_{\rm j}$ denote energy and velocity of the outflows, and $C_{\rm sj}~(= \sqrt{\Gamma P_{\rm j}/\rho_{\rm j}})$ is the sound speed of the outflowing matter. Here, $\Psi^{\rm eff}$ refers the effective potential and is given by \cite[]{Dihingia-etal2018a} ,
\begin{equation}
	\begin{split}
		\Psi^{\rm eff} & = \frac{1}{2} \ln \, \frac{x^2_{\rm j} (2 {\mathcal Y} r_{\rm j} - 4 a^2_{\rm k}x^2_{\rm j} - {\mathcal Y} {\mathcal Z})}{{\mathcal Z} [- {\mathcal Y} x^2_{\rm j} + 4 a_{\rm k} \lambda_{\rm j}r_{\rm j}x^2_{\rm j} + \lambda^2_{\rm j}r^2_{\rm j}({\mathcal Z} - 2 r_{\rm j})]},
	\end{split}
	\label{pot-off}
\end{equation}
where $\mathcal{Y}=(r^2_{\rm j} + a^2_{\rm k})^2- \Delta  a_{\rm k}^2 (x_{\rm j}/r_{\rm j})^2$, $\mathcal{Z} = { r^2_{\rm j} + a^2_{\rm k}\Big(1 - (x_{\rm j}/r_{\rm j})^2\Big)}$ and $r_{\rm j}~\left(=\sqrt{x^2_{\rm j} + z^2_{\rm j}}\right)$ is the spherical radius of outflow \cite[]{Jana-Das-2024}. In equation (\ref{outmass}), $\dot{M}_{\rm out}$ is the outflowing mass rate and $\mathcal{A}_{\rm j}$ is the area function. In this work, we adopt a jet geometry where the outflowing matter is confined between two boundaries, namely centrifugal wall (CB) and funnel wall (FW) \cite{Molteni-etal1996,Das-Chakrabarti2008,aktar-etal-2015,Jana-Das-2024}. The CB is defined as $\left(\frac{ d\Psi^{\rm eff}}{ dx}\right)_{r_{\rm CB}} = 0$ and the radius of funnel wall is determined using $\Psi^{\rm eff}\vert_{ r_{\rm FW}}$ = $0$. Subsequently, the jet coordinates are defined as, $x_{\rm j} = (x_{\rm CB} + x_{\rm FW})/2$ and $z_{\rm j} = z_{\rm CB} = z_{FW}$. With this, we calculate the area function as $\mathcal{A} = 2 \pi (x^2_{\rm CB} - x^2_{\rm FW})/\sqrt{1 + (dx_{\rm j}/dz_{\rm j})^2}$, where $\sqrt{1 + (dx_{\rm j}/dz_{\rm j})^2}$ represents the projection factor \cite[]{Kumar-Chattopadhyay-2013}. Similar to accretion, we carry out the critical point analysis and obtain the outflow velocity ($v_{\rm jc}$) at the outflow critical point ($r_{\rm jc}$) from the critical point conditions as \cite[]{das-chattopadhyay-2008, Jana-Das-2024},
\begin{equation}
    v_{\rm jc} = \sqrt{ \left(\frac{d \Psi^{\rm eff}}{dr}\right)_{r_{\rm jc}}
		\left[\frac{1}{\mathcal A_{\rm j}}\Big(\frac{ \mathcal{A}_{\rm j}}{dr}\Big)_{r_{\rm jc}}\right]^{-1} },
        \label{vjc}
\end{equation}
where the subscript `$\rm jc$' indicates the jet quantities at $r_{\rm jc}$ along with $r=r_{\rm CB}$. Using equation (\ref{vjc}), the outflow temperature at $r_{\rm jc}$ is determined by numerically solving the following equation:
\begin{equation}
    \sqrt{\frac{\Theta_{\rm jc}}{N} \Big[\frac{d }{d \Theta_{\rm j}}\left(N C^2_{\rm sj}\right)\Big]_{\Theta_{\rm jc}}} = v_{\rm jc}.
    \label{thetajc}
\end{equation}
 Employing $v_{\rm jc}$ and $\Theta_{\rm jc}$, we solve the jet equations (\ref{outeng}, \ref{outmass}) and uniquely obtain the outflow solutions for a given set of $\mathcal{E}_{\rm j}$ and $\lambda_{\rm j}$. In this formalism, outflows are considered to be originated from the PSC region, and hence, in the next section, we outline the methodology for solving the governing equations of accretion and outflows self-consistently incorporating Rankine-Hugoniot conditions \cite[]{Landau-Lifshitz-1959} for shock in inflowing matter.
 
\subsection{Disk-jet connection} 

During the accretion process, matter experiences two opposing forces, namely the gravitational pull and centrifugal repulsion. When these forces become comparable at the vicinity of the black hole, the supersonic inflow decelerates causing matter to accumulate. However, this accumulation does not continue indefinitely, instead, once a critical threshold is reached, centrifugal repulsion triggers a shock transition \cite[]{Frank-etal-2002}. The resulting shock compression makes the post-shock flow --- also known as the post-shock corona (PSC) --- dense and hot, forming a puffed-up, torus-like effective boundary around the black hole. The excess thermal gradient force across the shock front then drives a fraction of the infalling matter vertically outward as bipolar outflows. Since this ejection mechanism is linked to the accretion process through shock formation, we apply shock conditions to solve the inflow and outflow equations. In presence of mass loss, the shock conditions \cite[]{Landau-Lifshitz-1959, das-chattopadhyay-2008,Sarkar-etal2018} are as follows: (a) the mass conservation $\dot{M}_{+} = \dot{M}_{-} - \dot{M}_{\rm out} = \dot{M}_{-} (1 - R_{\dot{\rm m}})$, (b) the energy conservation: $\mathcal{E}_{+} = \mathcal{E}_{-}$, (c) the momentum conservation: $W_{+} + \Sigma_{+} v^2_{+} = W_{-} + \Sigma_{-} v^2_{-}$, and (d) the magnetic flux conservation $ \dot{\Phi}_{+} = \dot{\Phi}_{-}$. Here, $R_{\dot{\rm m}} ~(= \dot{M}_{\rm out}/\dot{M}_{-})$ is the outflow rate, and the symbols `$+$' and `$-$' refer the quantities immediately after and before the shock transition. 

 In this work, since outflows are assumed to be originated from PSC, we consider the  matter to eject with the same density as the post shock flow, $i.e$, $\rho_{\rm j} = \rho_{+}$. Based on this consideration, we compute the mass outflow rate $R_{\dot m}$ which is given by,
\begin{equation}
    R_{\dot{\rm m}} = \Big(\frac{\Sigma_{+}}{\Sigma_{-}}\Big) \frac{v_{\rm jb} \mathcal{A}_{\rm jb}}{4 \pi  v_{-} H_{+}\sqrt{\Delta}},
    \label{outflowrate}
\end{equation}
where $v_{\rm jb}$ and $\mathcal{A}_{\rm jb}$ represent the outflow speed and area function of outflow at the jet base (equivalently shock location), respectively. To solve the coupled inflow-outflow system self-consistently, we employ the successive iteration method \cite[]{chattopadhyay-Das-2007, das-chattopadhyay-2008}. We begin with $R_{\dot{m}} = 0$ and determine the virtual shock location ($\Tilde{x}_{\rm s}$), along with the corresponding inflow properties across the shock front. We use $\mathcal{E}_{\rm j} = \mathcal{E} (\Tilde{x}_{\rm s})$, $\lambda_{\rm j} = \lambda (\Tilde{x}_{\rm s})$ and compute the jet critical point ($r_{\rm jc}$) using equation (\ref{outeng}, \ref{vjc} and \ref{thetajc}), $v_{\rm jc}$ and $\Theta_{\rm jc}$. We employ the outflow variables at the critical point and solve equations \ref{outeng} and \ref{outmass} from $r_{\rm jc}$ towards BH, up to the shock location as it is considered as the jet base. Utilizing the inflow and outflow variables across the shock, we determined $\Tilde{R}_{\dot{\rm m}}$ using equation (\ref{outflowrate}). Next, we use $\Tilde{R}_{\dot{\rm m}}$ to calculate the updated shock location following the above method. We continue the iteration till the shock location converges and with this, we finally obtain $R_{\dot{\rm m}}$. We discuss the properties of $R_{\dot{\rm m}}$ in the presence of thermal conduction, in the subsequent sections.

\section{Results}
\label{section3}

\begin{figure}
    \includegraphics[width=\columnwidth]{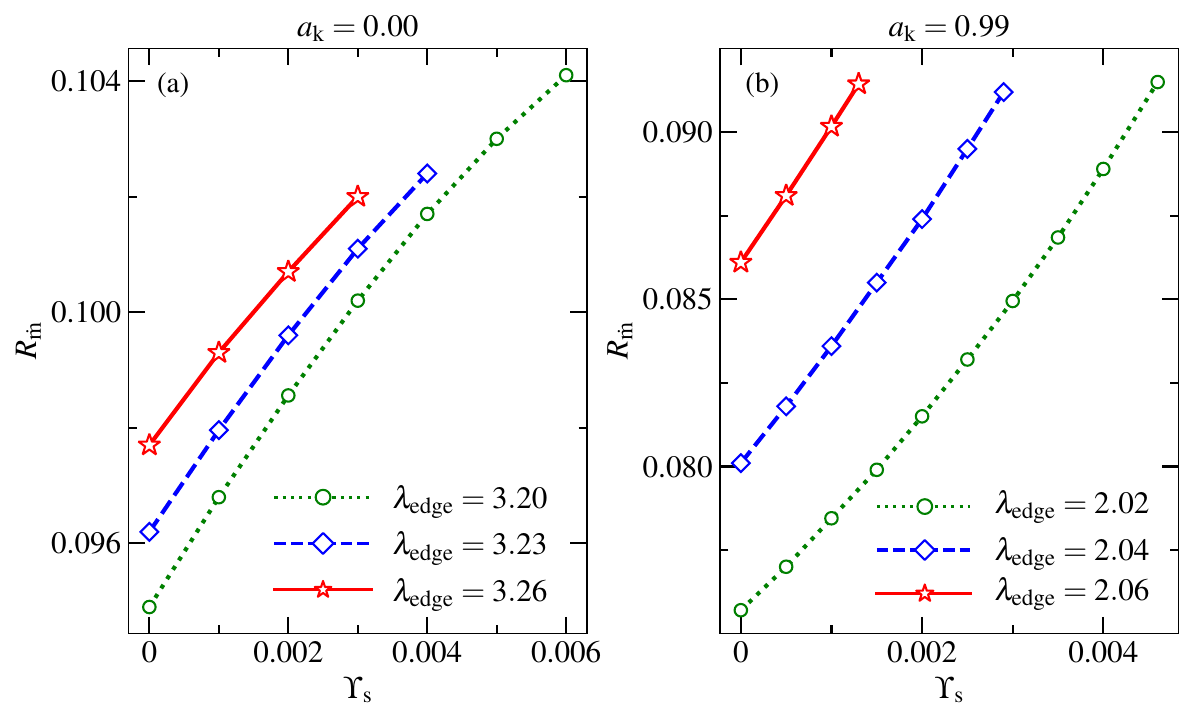}
    \caption{Variation of mass outflow rate ($R_{\dot{\rm m}}$) as a function of conduction parameter ($\Upsilon_{\rm s}$) for different $\lambda_{\rm edge}$ values, considering inflow injected from $x_{\rm edge} = 1000$. Panels (a) and (b) depict results for $a_{\rm k} = 0.0$ and $0.99$, respectively. In (a), we set $\alpha_{\rm B} = 0.001$, $\mathcal{E}{\rm edge} = 2 \times 10^{-4}$, and $\beta{\rm edge} = 10^5$, with $\lambda_{\rm edge}$ = $3.20$ (circle), $3.23$ (diamond), and $3.26$ (asterisk). In (b), $\alpha_{\rm B} = 0.001$, $\mathcal{E}_{\rm edge} = 1.2 \times 10^{-3}$, and $\lambda{\rm edge}$ = $2.02$ (circle), $2.04$ (diamond), and $2.06$ (asterisk). See the text for details.}    
    \label{lambda}
\end{figure}

We present our numerical results of mass outflow rate ($R_{\dot m}$) around both weakly rotating ($a_{\rm k} \rightarrow 0$) and rapidly rotating ($a_{\rm k}=0.99$) black holes. In Fig.\ref{lambda}, we depict the variation of $R_{\dot{\rm m}}$ with $\Upsilon_{\rm s}$ for flows with a fixed outer boundary. In Fig.\ref{lambda}a, we inject matter around non-rotating black hole ($a_{\rm k}=0$) from the outer edge of the disk at $x_{\rm edge} = 1000$ with $\mathcal{E}_{\rm edge} = 2.0 \times 10^{-4}$, $\alpha_{\rm B} = 0.001$ and $\beta_{\rm edge} = 10^5$. The dotted, dashed, and solid lines connecting the circles (green), diamonds (blue), and asterisks (red) are obtained for $\lambda_{\rm edge} = 3.20$, $3.23$, and $3.26$, respectively. We observe that for a given $\lambda_{\rm edge}$, $R_{\dot{\rm m}}$ increases with the increase of $\Upsilon_{\rm s}$, which is consistent with the prediction of \cite{Tanaka-Menou-2006, Rezgui-etal-2019}. Indeed, for a set of inflow parameters, increasing $\Upsilon_{\rm s}$ boosts the thermal pressure that eventually pushes the shock front outwards, thereby inflating the size of PSC \cite{Singh-Das-2024b}. As a result, the jet launching area increases. In addition, increased thermal conduction raises the local energy ($\mathcal{E}$) of the inflowing matter (see Fig \ref{app1} in Appendix B), which in turn increases the flow energy across the shock front ($x_{\rm s}$). Overall, a higher $\Upsilon_{\rm s}$ enhances the jet energy, providing a stronger driving force that ejects matter at a higher velocity leading to an enhanced outflow rate ($R_{\dot{\rm m}}$). On contrary, for a fixed $\Upsilon_{\rm s}$, $R_{\dot{\rm m}}$ increases with $\lambda_{\rm edge}$. This finding is not surprising because higher inflow angular momentum yields stronger centrifugal barrier, causing the shock to form at a larger radius, and hence, the size of PSC increases. Consequently, more of the inflowing material is deflected at the enhanced jet base, leading to an increase in the outflow rate ($R_{\dot{\rm m}}$). Next, in Fig.\ref{lambda}b, we choose $\mathcal{E}_{\rm edge} = 1.2 \times 10^{-3}$, $\alpha_{\rm B} = 0.001$, $\beta_{\rm edge} = 10^5$ and inject matter around rapidly rotating black hole ($a_{\rm k}=0.99$) with different $\lambda_{\rm edge}$. Circles (green), diamonds (blue) and asterisks (red) joined with dotted, dashed and solid lines represent results for  $\lambda_{\rm edge}=2.02$, $2.04$, and $2.06$, respectively. In general, we observe a strong correlation between $R_{\dot{\rm m}}$ and $\Upsilon_{\rm s}$ when $\lambda_{\rm edge}$ held fixed. Moreover, similar to the case of a weakly rotating black hole, $ R_{\dot{\rm m}} $ remains higher for larger $ \lambda_{\rm edge} $ when $ \Upsilon_{\rm s} $ is fixed. Interestingly, outflows are generated from the rapidly rotating black hole even when $\lambda_{\rm edge}$ is relatively smaller compared to that of a weakly rotating black hole. This occurs primarily due to the spin-orbit coupling present in the effective potential of the black hole spacetime, where the marginally stable angular momentum strongly anti-correlates with the black hole spin $a_{\rm k}$. Furthermore, we observe that for a fixed $\lambda_{\rm edge}$, mass loss no longer occurs beyond a limiting value of $\Upsilon_{\rm s}$ (say $\Upsilon^{\rm max}_{\rm s}$), as RHCs are not favorable in that regime. Moreover, $\Upsilon^{\rm max}_{\rm s}$ decreases with $\lambda_{\rm edge}$ irrespective to $a_{\rm k}$ values. It is important to note that $\Upsilon^{\rm max}_{\rm s}$ does not assume universal values, as it primarily depends on the inflow parameters.

\begin{figure}
    \includegraphics[width=\columnwidth]{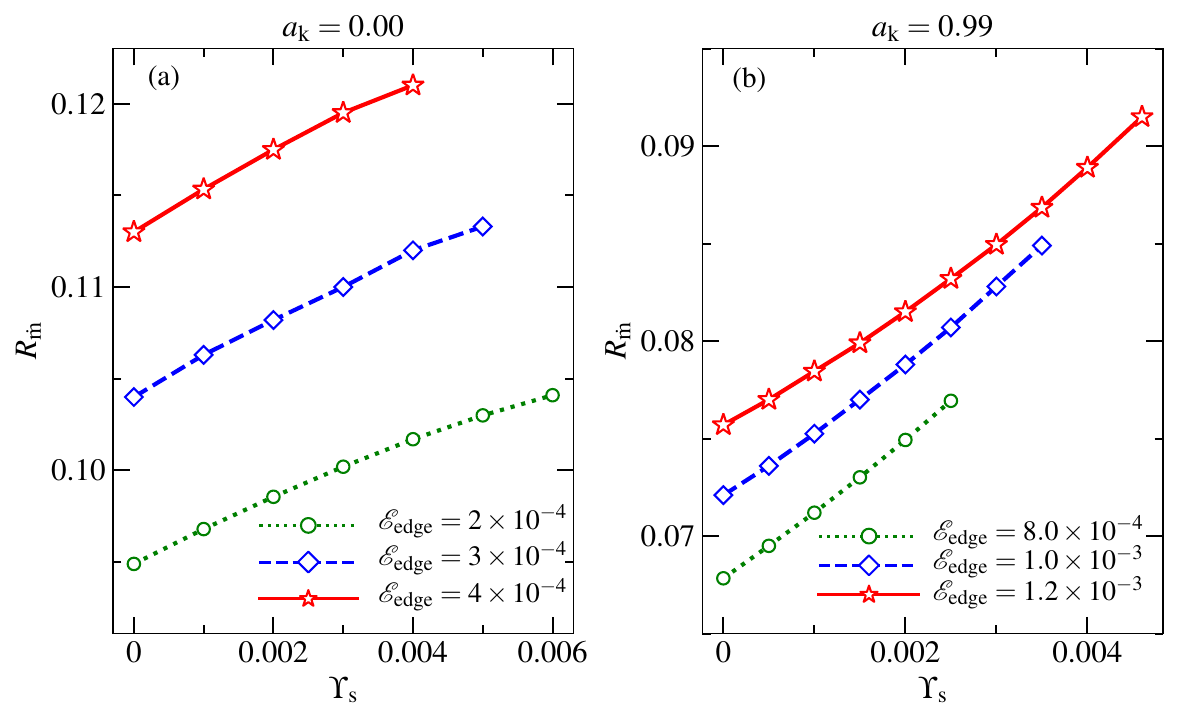}
    \caption{Variation of $R_{\dot{\rm m}}$ with $\Upsilon_{\rm s}$ for a set of inflow energy $\mathcal{E}_{\rm edge}$. In panel (a), we choose $a_{\rm k} = 0.0$, and inject matter from $x_{\rm edge} = 1000$ with $\lambda_{\rm edge} = 3.20$ and $\beta_{\rm edge} = 10^5$ and $\alpha_{\rm B} = 0.001$. Circles, diamonds and asterisks joined with dotted (green), dashed (blue) and solid (red) lines denote results for $\mathcal{E}_{\rm edge} = 0.0002, 0.0003$, and $0.0004$, respectively. In panel (b), we set $a_{\rm k} = 0.99$ and $\alpha_{\rm B} = 0.001$, and fix $\lambda_{\rm edge} = 2.02$, and $\beta_{\rm edge} = 10^{5}$ at $x_{\rm edge} = 1000$. Circles, diamonds, and asterisks connected with dotted (green), dashed (blue), and solid (red) lines represent result for $\mathcal{E}_{\rm edge}$ = $8.0 \times 10^{-4}$, $1.0 \times 10^{-3}$, and $1.2 \times 10^{-3}$, respectively. See the text for details.}
    \label{eng}
\end{figure}

Next, we put effort to understand the influence of inflow energy in determining the mass outflow rate in presence of thermal conduction. In doing so, we examine how $R_{\dot{\rm m}}$ varies with $\Upsilon_{\rm s}$ for varied inflow energies. Here, we inject inflowing matter from the outer boundary at $x_{\rm edge}=10^3$ with different energy values ($\mathcal{E}_{\rm edge}$). The obtained results are shown in Fig. \ref{eng}. In Fig. \ref{eng}a, we present the results for non-rotating black hole with inflow parameters set to $\alpha_{\rm B} = 0.001$, $\lambda_{\rm edge} = 3.20$, and $\beta_{\rm edge} = 10^5$. Dotted, dashed, and solid lines joining circles (green), diamonds (blue) and asterisks (red) correspond to the results obtained for $\mathcal{E}_{\rm edge} = 2\times 10^{-4}$, $3\times 10^{-4}$, and $4\times 10^{-4}$, respectively. Similarly, in Fig. \ref{eng}b, we present the variation of $R_{\dot{\rm m}}$ with $\Upsilon_{\rm s}$ around rapidly rotating black hole of spin $a_{\rm k}=0.99$. Here, we choose $\alpha_{\rm B} = 0.001$, $\lambda_{\rm edge} = 2.02$, and $\beta_{\rm edge} = 10^5$. Results obtained for $\mathcal{E}_{\rm edge} = 8.0\times 10^{-4}$, $1.0\times 10^{-3}$, and $1.2\times 10^{-3}$ are plotted using circles, diamonds and asterisks joined with dotted, dashed and solid lines. In both panels, we find that for a fixed $\Upsilon_{\rm s}$, $R_{\dot{\rm m}}$ increases with $\mathcal{E}_{\rm edge}$. This is expected, as higher $\mathcal{E}_{\rm edge}$ leads to the higher jet driving force resulting in an increased outflow rate.

\begin{figure}
    \centering
     \includegraphics[width=\columnwidth]{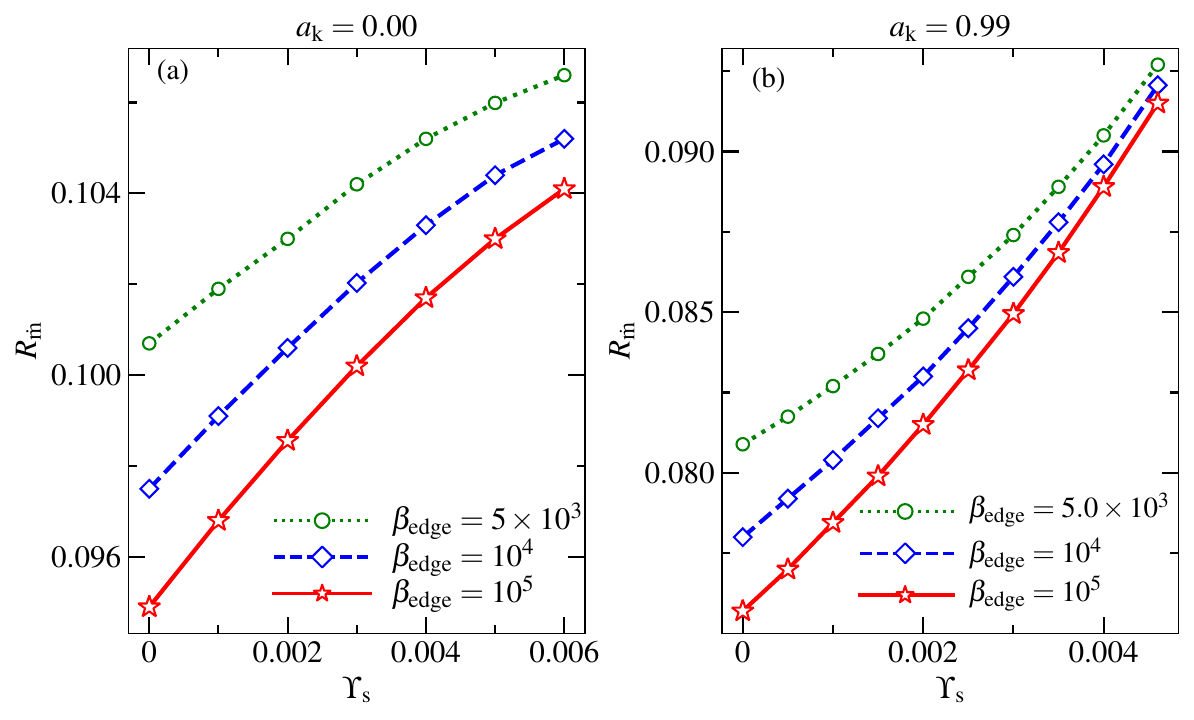}
     \caption{Variation of $R_{\dot{\rm m}}$ as a function of $\Upsilon_{\rm s}$ for a set of plasma-$\beta_{\rm edge}$ parameter. Panel (a) corresponds to BH spin parameter $a_{\rm k} = 0.0$. Here, we choose $\alpha_{\rm B} = 0.001$ and the other input parameters fixed at $x_{\rm edge} = 1000$ are $\lambda_{\rm edge} = 3.20$ and $\mathcal{E}_{\rm edge} = 2 \times 10^{-4}$. Circles, diamonds, and asterisks joined with dotted (green), dashed (blue), and solid (red) lines represent results obtained for $\beta_{\rm edge} = 5 \times 10^3$, $ 10^4$ and $10^5$, respectively. In panel (b), we set $a_{\rm k} = 0.99$ and $\alpha_{\rm B} = 0.001$, and inject matter with $\lambda_{\rm edgee} = 2.02$, and $\mathcal{E}_{\rm edge} = 1.2 \times 10^{-3}$ from $x_{\rm edge}=10^3$. Circles, diamonds, and asterisks joined with dotted (green), dashed (blue), and solid (red) lines are for $\beta_{\rm edge}$ = $5 \times 10^3$, $10^4$, and $10^5$, respectively. See the text for details. }
    \label{beta}
\end{figure}

We continue our investigation on mass loss as it is intriguing to explore the role of magnetic fields in generating outflows from the accretion disk in the presence of thermal conduction. To this end, we estimate $ R_{\dot{\rm m}} $ by varying the plasma $ \beta$ of the inflowing matter. The obtained results are shown in Fig. \ref{beta}, where the variation of $R_{\dot{\rm m}}$ is depicted as a function of $\Upsilon_{\rm s}$ for a set of $\beta_{\rm edge}$ values. In Fig. \ref{beta}a, we present the results for a non-rotating black hole ($a_{\rm k} = 0.0$), where the inflowing matter is injected from the outer edge of the disk at $x_{\rm edge} = 1000$ with $\alpha_{\rm B} = 0.001 $, $ \lambda_{\rm edge} = 3.20 $, and $ \mathcal{E}_{\rm edge} = 2 \times 10^{-4}$. The circles, diamonds, and asterisks connected by the dotted (green), dashed (blue), and solid (red) lines denote results obtained for $\beta_{\rm edge} = 5 \times 10^3$, $ 10^4$, and $ 10^5$, respectively. Similarly, in Fig. \ref{beta}b, we set $a_{\rm k} = 0.99$ and fix $ \alpha_{\rm B} = 0.001$, $\lambda_{\rm edge} = 2.02$, and $ \mathcal{E}_{\rm edge} = 1.2 \times 10^{-3} $ for inflowing matter injected from $x_{\rm edge} = 1000$. The circles, diamonds, and asterisks joined using dotted (green), dashed (blue), and solid (red) lines are for $\beta_{\rm edge}$ = $5 \times 10^3$, $10^4$, and $10^5$, respectively. From the figure, it is evident that $ R_{\dot{\rm m}} $ generally increases with $\Upsilon_{\rm s}$ regardless of the strength of the disk magnetic fields ($\beta_{\rm edge}$) for both non-rotating and rapidly rotating black hole cases. Moreover, for a fixed value of $\Upsilon_{\rm s}$, a higher $\beta_{\rm edge}$ leads to an increased mass loss from the magnetized disk. As the magnetic field strength increases, it enhances angular momentum transport, causing the shock to settle down at a smaller radius. Because of this, the inflow energy at PSC increases that strengthens the jet-driving leading to more mass loss. It is worth noting that for rapidly rotating black hole ($a_{\rm k}=0.99$), the influence of the plasma-$\beta$ parameter on $R_{\dot{\rm m}}$ is more pronounced at relatively lower $\Upsilon_{\rm s}$ values, while $R_{\dot{\rm m}}$ tends converge at limiting $\Upsilon_{\rm s}$ values.

Overall, we find that $R_{\dot{\rm m}}$ dynamically scales with $\Upsilon_{\rm s}$ across inflows with different $\mathcal{E}_{\rm edge}$, $\lambda_{\rm edge}$, and $\beta_{\rm edge}$. These results clearly highlight the pivotal role of energy, angular momentum, and magnetic fields in governing mass loss from the magnetized disk around both weakly as well as rapidly rotating black holes.

\begin{figure}
    \centering
    \includegraphics[width=\columnwidth]{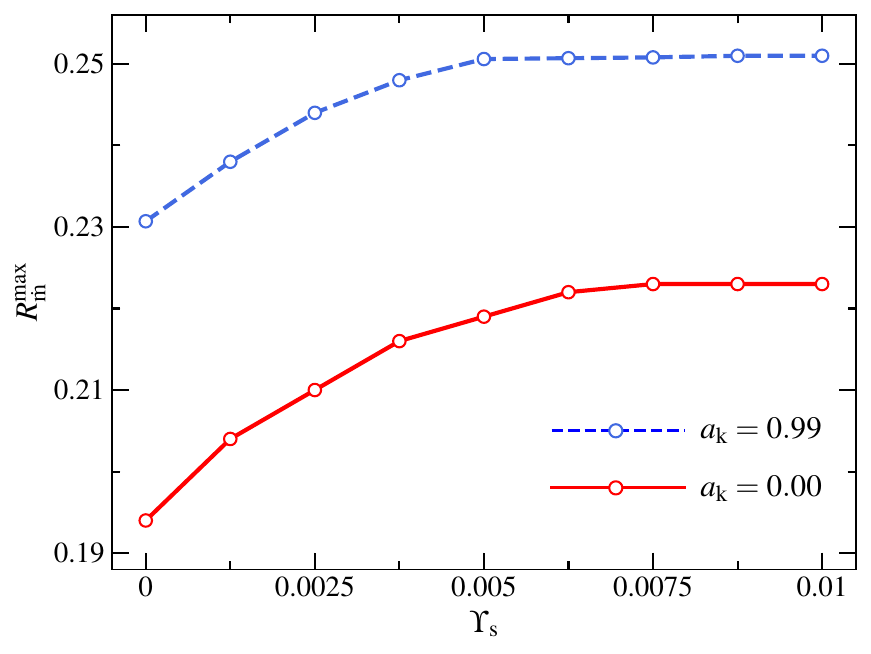}
    \caption{Variation of maximum outflow rate $R^{\rm max}_{\dot{\rm m}}$ with the conduction parameter $\Upsilon_{\rm s}$. Circles joined using solid (red) and dashed (blue) lines denote results corresponding to $a_{\rm k} = 0.0$ and $0.99$, respectively. Here, we choose $\beta_{\rm in} = 100$, and $\alpha_{\rm B} = 0.01$. See the text for details.}
    \label{maxr}
\end{figure}

Finally, we put efforts to estimate the maximum mass outflow rate ($R^{\rm max}_{\dot{\rm m}}$) from the magnetized accretion disk. To achieve this, we compute $R_{\dot{\rm m}}^{\rm max}$ as a function of $ \Upsilon_{\rm s}$ for two extreme cases of BH spin: a non-rotating black hole ($a_{\rm k} = 0.0$) and a rapidly rotating black hole ($a_{\rm k} = 0.99$). Here, we set $\alpha_{\rm B} = 0.01$ and fix the plasma-$\beta$ at the inner critical point ($x_{\rm in}$) as $\beta_{\rm in} = 100$, while allowing the energy ($\mathcal{E}_{\rm in}$) and angular momentum ($\lambda_{\rm in}$) at $x_{\rm in}$ to vary freely. The obtained results are presented in Fig. \ref{maxr}, where circles connected by solid (red) and dashed (blue) lines representing results for non-rotating ($a_{\rm k} = 0.0$) and rapidly rotating ($a_{\rm k} = 0.99$) black holes, respectively. Our findings clearly reveal that higher $a_{\rm k}$ drives stronger outflows regardless of the conduction parameter ($\Upsilon_{\rm s}$), which remain consistent with the results reported earlier in weak thermal conduction limit \cite[]{Jana-Das-2024}. Interestingly, for a fixed $a_{\rm k}$, $R_{\dot{\rm m}}^{\rm max}$ initially increases with $\Upsilon_{\rm s}$, however eventually tends to saturate. Furthermore, we note that for a rapidly rotating BH with $a_{\rm k} = 0.99$, $R_{\dot{\rm m}}^{\rm max}$ peaks at an impressive $25\%$. These results indeed suggest that accretion flows around rapidly spinning black holes with higher thermal conduction are significantly more susceptible to mass loss.

Until now, we have explored how inflow parameters ($\Upsilon, \mathcal{E}, \lambda, \beta$) and black hole spin ($a_{\rm k}$) influence the outflow dynamics. Now, it is intriguing to examine how this theoretical framework connects to the powerful radio jets frequently observed in active galactic nuclei. In the following section, we extend this accretion-ejection formalism to quantify radio jet power using the mass outflow rate ($R_{\dot{\rm m}}$) for low-luminosity AGNs (LLAGNs).

\section{Astrophysical Implications}
\label{section4}

In this section, we compare the predictions of our theoretical model with the observational findings. Our present formalism is based on the premise that the disk is radiatively inefficient and shows jet activity. Thus, this model can be suitable for explaining the radio loud low luminosity AGNs (LLAGNs). To start with, we follow the LLAGNs catalog from \cite{Liu-etal-2019} (hereafter L19), which includes total $14,585$ AGNs observed from SDSS DR7 at redshift factor, $z' < 0.35$ and BH masses ranging from $10^{5.1}-10^{10.3 }M_{\odot}$. To identify the radio counterpart of the sources, we cross-match the L19 catalogue with $1.4$ GHz FIRST survey \cite[]{White-etal-1997} using a search radius of $2$ arcsec. Subsequently, we find that 11.7 percent AGNs of the L19 catalog were detected in radio frequencies. Thereafter, to identify the radio loud AGNs, we compute radio loudness parameter $R$, which is defined as the ratio of $1.4$ GHz radio flux to the optical g-band flux, and set the condition $R >19$ \cite[]{Komossa-etal-2006} for categorizing the sources as radio loud. Based on these criteria, we finally select a sample of 68 radio-loud LLAGNs that exhibit quasi-simultaneous X-ray and radio observations.

Using the radio flux ($F_{\nu}$, $\nu$ is the radio frequency) provided by FIRST catalog at $1.4$ GHz, we estimate the radio luminosity $L_{1.4}$ (in Watt $\rm Hz^{-1}$) as,
\begin{equation}
    L_{1.4} = 4 \pi \times 10^{-7} \frac{D^2_{\rm L}}{(1 + z')^{1 + \alpha}} \times F_{1.4},
\end{equation}
where $D_{\rm L}$ represents the luminosity distance and $\alpha$ is the spectral index. Here, we set $\alpha = - 0.8$ \cite[]{Condon-1992} considering $F_{\nu} \propto \nu^{\alpha}$ for the computation. Next, we estimate the radio luminosity $L_{\rm R}$ (in $\rm erg \, s^{-1}$) at $5$ GHz utilizing $L_{1.4}$ in the adopted relation given by \cite[]{Yuan-etal-2018},
\begin{equation}
    \log L_{\rm R} = (20.9 \pm 2.1) + (0.77 \pm 0.08) \log L_{1.4}.
\end{equation}
We observe that $L_{\rm R}$ varies within $10^{36.91} - 10^{39.66}$ $\rm erg \, s^{-1}$ for selected AGNs under considerations. Following \cite[]{Blandford-Konigl1979,Falcke-Biermann-1995,Heinz-Sunyaev2003}, we estimate the jet power ($L^{\rm obs}_{\rm jet}$) from the observed radio luminosity $L_R$ as
\begin{equation}
    L^{\rm obs}_{\rm jet} \propto L_{R}^{12/17}.
\end{equation}
Meanwhile, \cite[]{Heinz-Grimm-2005} established the relationship between the jet power and the observed jet luminosity, which is given by,
\begin{equation}
    L^{\rm obs}_{\rm jet} = \mathcal{W}_{0} \Big(\frac{L_{\rm R}}{L_{0}}\Big)^{12/17},
\end{equation}
where $L_0 = 1.6 \times 10^{30}$ $\rm erg \, s^{-1}$ and $\mathcal{W}_{0}$ is the normalization of the kinetic jet power to observed radio luminosity. Considering $\mathcal{W}_{0} \sim 6.2 \times 10^{37}$ $\rm erg \, s^{-1}$ for AGNs  \cite[]{Heinz-Grimm-2005}, we coarsely express the jet power as,
\begin{equation}
    L^{\rm obs}_{\rm jet} = 2.96 \times 10^{16} L_{\rm R}^{12/17}.
    \label{norm}
\end{equation}

Further, X-ray luminosity ($L_{\rm x}$) for the same sample of AGNs is calculated in $0.2 - 12$ keV energy range from XMM$-$Newton data \cite[]{Rosen-etal-2016}, which spans in the range of $10^{40.99} - 10^{45.02}$ $\rm erg \, s^{-1}$. Using $L_{\rm x}$, we estimate the accretion power as $\dot{M}c^2=L_{\rm x}/\eta_{\rm acc}$, where $\eta_{\rm acc}$ denotes the accretion efficiency factor. Since the spin measurements of supermassive black holes (SMBHs) in these AGNs are not well constrained, we adopt a maximum efficiency of $\eta_{\rm acc} = 0.3$ in this study \cite[]{Throne-1974}. Thereafter, we employ the accretion-ejection model formalism to calculate the maximum kinetic jet power \cite[]{aktar-etal-2015,Nandi-etal-2018,Jana-Das-2024} using maximum mass outflow rate $R^{\rm max}_{\dot m}$ as
\begin{equation}
    L^{\rm max}_{\rm jet} =  R^{\rm max}_{\dot{\rm m}} \times \dot{M} \times c^2 {~\rm erg ~s}^{-1}.
\end{equation}

\begin{figure}
    \centering
    \includegraphics[width=\columnwidth]{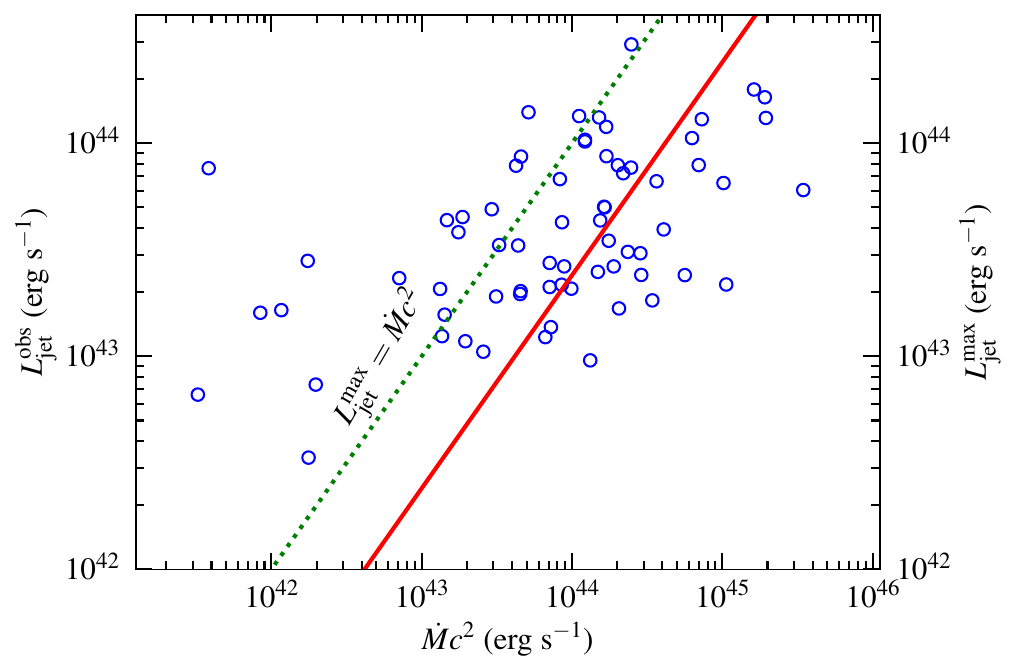}
    \caption{Plot of observed jet power $L^{\rm obs}_{\rm jet}$ and theoretically obtained maximum jet kinetic power $L^{\rm max}_{\rm jet}$ with accretion power (${\dot M}c^2$). Open circles denote $L^{\rm obs}_{\rm jet}$ for LLAGNs and solid (red) line refers the model estimated $L^{\rm max}_{\rm jet}$. Dotted (green) line represents equality jet power and accretion power $L^{\rm max}_{\rm jet}={\dot M}c^2$. See the text for details.
   }
    \label{ljet}
\end{figure}

It is indeed evident that the maximum outflow rate depends on the black hole spin, and the present formalism yields $R^{\rm max}_{\dot{\rm m}} \simeq 0.25 ~(0.22)$  for  $a_{\rm k} = 0.99~(0.0)$ (see Fig. \ref{maxr}). Therefore, to compute $L^{\rm max}_{\rm jet}$ in terms of the accretion power, we adopt $R^{\rm max}_{\dot{\rm m}} \simeq 0.25$ for the purpose of representation without loosing the generality. Subsequently, in Fig. \ref{ljet}, we compare the observed jet power ($L^{\rm obs}_{\rm jet}$) with the theoretically estimated maximum jet kinetic power $L^{\rm max}_{\rm jet}$. The open circles represent the observed jet power for the radio-loud LLAGNs under consideration, while the solid (red) line indicates the upper limit of the theoretical prediction for jet kinetic power. It is evident that a substantial number of LLAGNs fall below the theoretical limit, suggesting that the current accretion-ejection model formalism effectively accounts for the jet kinetic power.

It is important to note that in this work, jets are assumed to originate from the accretion disk, making the model particularly applicable when the jet power is less than the accretion power, $i.e.$, $L^{\rm max}_{\rm jet}/\dot{M} c^2 < 1$. However, a few sources are observed to exceed the line $L^{\rm max}_{\rm jet} = \dot{M} c^2$ (dotted line in green), suggesting that their jet power surpasses the accretion power. Here, we note that estimation of these observables involve significant uncertainty due to various unknown factors affecting their calculation. For these sources, the additional jet energy is likely to be extracted from the black hole's rotational energy via the Blandford-Znajek (BZ) mechanism \cite[]{Blandford-Znajek-1977}. Overall, with the exception of these LLAGNs, our accretion-ejection model seems to provide a satisfactory explanation for the observed jet power of LLAGNs under consideration.
  
\section{SUMMARY and CONCLUSION}
\label{section5}

In this study, we examine the effect of thermal conduction on the mass loss from a low angular momentum, advective, viscous, and magnetized accretion disk around a rotating black hole, for the first time to the best of our knowledge. In doing so,  we assume the accreting plasma is collisionless and adopt the saturated form of thermal conduction \cite[]{Cowie-Mckee-1977}. In addition, we model the accretion disk as being threaded with a dominant toroidal magnetic field component \cite[]{Oda-etal-2007, Sarkar-Das2016, Sarkar-etal2018, Jana-Das-2024}. To simplify the general relativistic complexities, we adopt a recently developed effective potential \cite[]{Dihingia-etal2018a} that satisfactorily mimics the space-time around a rotating black hole. Furthermore, we employ the relativistic equation of state (REoS) \cite[]{Chattopadhyay-Ryu-2009} to determine the thermodynamic quantities. With these assumptions, we solve the coupled governing equations for both inflow and outflow self-consistently in terms of the inflow parameters, namely energy, angular momentum, plasma-$\beta$, and the conduction parameter. We summarize the key findings of this study below:

\begin{itemize}
    \item Using the global inflow-outflow solution, we compute the mass outflow rate ($R_{\dot {\rm m}}$) from the magnetized disk in presence of thermal conduction. We find that mass loss continues to happen for wide range of the inflow parameters for both weakly rotating ($a_{\rm k} \rightarrow 0 $) and rapidly rotating ($a_{\rm k} \rightarrow 0.99$) black holes.
    
    \item We observe that thermal conduction enhances mass loss from the magnetized disk, with $R_{\dot{\rm m}}$ increasing as the conduction parameter $\Upsilon_{\rm s}$ rises, irrespective of the black hole spin values. We also notice that for a fixed $\Upsilon_{\rm s}$, $R_{\dot{\rm m}}$ increases as the energy, angular momentum, and magnetic field strength of the inflowing matter increase. However, mass loss ceases beyond a limiting value of $\Upsilon_{\rm s}$ as RHCs for shocks in inflowing matter becomes unfavorable (see Fig. \ref{lambda},\ref{eng} and \ref{beta}).
    
    \item We compute the maximum mass outflow rate $ R^{\rm max}_{\dot{\rm m}}$ as a function of $\Upsilon_{\rm s}$ for two extreme black hole spin parameters, $a_{\rm k} = 0.0$ and $0.99$ (Fig. \ref{maxr}). Our results reveal that rapidly rotating black holes expel more mass from the magnetized disk in the form of outflows compared to the weakly rotating black holes. More precisely, when thermal conduction is active within the disk, $R_{\dot{\rm m}}^{\rm max}$ reaches up to $25\%$ for $a_{\rm k} = 0.99$, whereas it remains below $22\%$ for $a_{\rm k} = 0.0$ (see Fig. \ref{maxr}).
    
    \item Finally, we investigate the implications of our accretion-ejection formalism in explaining the jet luminosity ($L^{\rm obs}_{\rm jet}$) observed from low-luminosity active galactic nuclei (LLAGNs). Towards this, we compare $L^{\rm obs}_{\rm jet}$ for $68$ radio-loud LLAGNs with the model-predicted maximum jet kinetic power ($L^{\rm max}_{\rm jet}$), and good agreement is seen for a significant number of sources (see Fig. \ref{ljet}).
    
\end{itemize}

It is important to mention that the present accretion-ejection formalism is developed based on few assumptions and approximations. We adopt an effective potential to describe the spacetime geometry of the black hole instead of full-fledged general relativistic calculations. Moreover, our model considers only the azimuthal component of the magnetic field in the accretion flow, omitting the poloidal component and neglecting magnetic fields in the outflow dynamics. Indeed, all these aspects seem to be relevant in the context of jet generation, however, their implementation lies beyond the scope of this study, and we intend to explore them in future endeavor.

\section*{Data Availability}

The data underlying this paper will be available with reasonable request.

\section*{Acknowledgements}

Authors thank the Department of Physics, IIT Guwahati, India for providing the infrastructural support to carry out this work.

\appendix

\section*{Appendix A: Derivation of the inflow governing equations}

Using equations (\ref{inmass}) and (\ref{eos}) in equations (\ref{ineng}), (\ref{inlam}), (\ref{entro}), and (\ref{mgflux}), we get
\begin{subequations}
    \begin{equation}
        R_{\rm v} \frac{dv}{dx} + R_{\Theta} \frac{d\Theta}{dx} + R_{\lambda} \frac{d\lambda}{dx} + R_{\beta} \frac{d\beta}{dx} + R_{0} = 0,
    \end{equation}

    \begin{equation}
        L_{\rm v} \frac{dv}{dx} + L_{\Theta} \frac{d\Theta}{dx} + L_{\lambda} \frac{d\lambda}{dx} + L_{\beta} \frac{d\beta}{dx} + L_{0} = 0,
    \end{equation}

    \begin{equation}
        E_{\rm v} \frac{dv}{dx} + E_{\Theta} \frac{d\Theta}{dx} + E_{\lambda} \frac{d\lambda}{dx} + E_{\beta} \frac{d\beta}{dx} + E_{0} = 0,
    \end{equation}

    \begin{equation}
        B_{\rm v} \frac{dv}{dx} + B_{\Theta} \frac{d\Theta}{dx} + B_{\lambda} \frac{d\lambda}{dx} + B_{\beta} \frac{d\beta}{dx} + B_{0} = 0,
    \end{equation}
\end{subequations}

The coefficients are given as 
\begin{align*}
    R_{\rm v} &= v + T_{\rm v},\quad R_{\Theta} = T_{\Theta}, \quad R_{\lambda} = T_{\lambda},\\ 
    R_{\beta} &= T_{\beta}, \quad {\rm and} \quad R_{0} = T_{0} + 2 t \Theta/(\beta x) + d\Psi_{\rm e}^{\rm eff}/dx,
\end{align*}
where 
\begin{align*}
t &= 2/\tau, T_{\rm v} = - \frac{t \Theta(\beta + 1)}{v\beta}, T_{\Theta} = \frac{t (\beta+ 1)}{2 \beta},\\
T_{\lambda} &= - \left(\frac{\beta + 1}{\beta }\right) \left(\frac{t \Theta}{2 \mathcal{F}} \right)\frac{\partial \mathcal{F}}{\partial \lambda}, T_{\beta} = - \frac{t \Theta}{2 \beta^2}, \\
T_{0} &= - t \Theta \left(\frac{\beta + 1}{\beta }\right) \left(\frac{1}{2 \Delta} \frac{d \Delta}{dx}\right) + \frac{1}{2 \mathcal{F}} \frac{\partial \mathcal{F}}{\partial x}.
\end{align*}

\begin{align*}
    L_{\rm v} &= \frac{\alpha_{\rm B} x \Theta t}{v} \Big(\frac{\beta + 1}{\beta}\Big) - \alpha_{\rm B} x v, \\
    L_{\Theta} &= - \alpha_{\rm B} x t \Big(\frac{\beta + 1}{\beta}\Big), \\
     L_{\lambda} &= v, \quad L_{\beta} = \frac{\alpha_{\rm B} x }{\beta^2} \Theta t, \\
     L_{0} & = \frac{\alpha_{\rm B}}{2 \beta \Delta}\Big(\frac{d \Delta}{dx} x - 4 \Delta \Big) \Big(v^2 \beta + (1 + \beta) \Theta t \Big).\\
    E_{\rm v} &= t \Theta + \frac{5 \Upsilon_{\rm s} (t \Theta)^{3/2}}{v},\\
    \quad E_{\Theta} &= (vt/2 + Nvt) - 5 \Upsilon_{ \rm s} t^{3/2} \sqrt{\Theta}, \\
     E_{\lambda} &= \frac{v t \Theta}{2 \mathcal{F}}\frac{\partial \mathcal{F}}{\partial \lambda} - \alpha_{\rm B} \Big(v^2 + \Theta t \frac{\beta + 1}{\beta }\Big) x \frac{\partial \Omega}{\partial \lambda} \\
     &+ \frac{5 \Upsilon_{\rm s} (t \Theta)^{3/2}}{2} \frac{1}{\mathcal{F}}\frac{\partial \mathcal{F}}{\partial \lambda},\\    
    E_{\beta} & = - \frac{v t \Theta}{2 (1 + \beta)\beta} - \frac{5 \Upsilon_{ \rm s} (t \Theta)^{3/2}}{2 \beta (1 + \beta)},\\
    E_{0} & = v t \Theta \Big(\frac{1}{2 \Delta} \frac{d \Delta}{dx} + \frac{1}{2 \mathcal{F}} \frac{\partial \mathcal{F}}{\partial x}\Big) \\
    &- \alpha_{\rm B} \Big(v^2 + \Theta t \frac{\beta + 1}{\beta }\Big) x \frac{\partial \Omega}{\partial x}\\
    & + 5 \Upsilon_{\rm s} (t \Theta)^{3/2} \Big(- \frac{1}{x} + \frac{1}{2 \Delta} \frac{d \Delta}{dx} + \frac{1}{2 \mathcal{F}} \frac{\partial \mathcal{F}}{\partial x}\Big).
\end{align*}

\begin{align*}
    B_{\rm v} &= \frac{1}{2 v}, \quad B_{\Theta} = \frac{3}{4 \Theta}, \quad B_{\lambda} = \frac{1}{4 \mathcal{F}} \frac{\partial \mathcal{F}}{\partial \lambda}\\
    B_{\beta} & = - \frac{1}{4 \beta (1 + \beta)} - \frac{1}{2 \beta}, \quad {\rm and}\\
    B_{0} & = \frac{1}{4} \Big(  \frac{1}{ \mathcal{F}} \frac{\partial \mathcal{F}}{\partial x} - \frac{1}{ \Delta} \frac{d \Delta}{dx})\Big) + \frac{ 1}{x}.
\end{align*}

\begin{align*}
    N_{\Theta} = &R_{\Theta} B_{\beta} L_{\lambda} - R_{\Theta} B_{\lambda} L_{\beta} - R_{\lambda} B_{\beta} L_{\Theta} + \\
    & R_{\beta} B_{\lambda} L_{\Theta} - R_{\beta} B_{\Theta} L_{\lambda} + R_{\lambda} B_{\Theta} L_{\beta},\\
    N_{\rm v} = &R_{\rm v} B_{\lambda} L_{\beta} + R_{\lambda} B_{\beta} L_{\rm v} + R_{\beta} B_{\rm v} L_{\lambda} - \\
    & R_{\rm v} B_{\beta} L_{\lambda} - R_{\beta} B_{\lambda} L_{\rm v} - R_{\lambda} B_{\rm v} L_{\beta},\\
    N_{0} = &R_{0} B_{\lambda} L_{\beta} + R_{\lambda} B_{\beta} L_{0} + R_{\beta} B_{0} L_{\lambda} - \\
    & R_{0} B_{\beta} L_{\lambda} - R_{\beta} B_{\lambda} L_{0} - R_{\lambda} B_{0} L_{\beta}.
\end{align*}

\begin{equation*}
    \Theta_{\rm v} = \Big(\frac{N_{\rm v}}{N_{\Theta}}\Big), \quad \Theta_{0} = \Big(\frac{N_{0}}{N_{\Theta}}\Big).
\end{equation*}

\begin{align*}
    b_{0} &= \frac{B_{\lambda}L_{0} - B_{0}L_{\lambda}}{B_{\beta} L_{\lambda} - B_{\lambda}L_{\beta}}, \  b_{\rm v} = \frac{B_{\lambda}L_{\rm v} - B_{\rm v}L_{\lambda}}{B_{\beta} L_{\lambda} - B_{\lambda}L_{\beta}},\\
      b_{\Theta} &= \frac{B_{\lambda}L_{\Theta} - B_{\Theta}L_{\lambda}}{B_{\beta} L_{\lambda} - B_{\lambda}L_{\beta}}
\end{align*}

\begin{align*}
    \beta_{\rm v} = b_{\Theta} \Theta_{\rm v} + b_{\rm v}, \quad \beta_{0} = b_{\Theta} \Theta_{0} + b_{0}.\\
    \lambda_{\rm v} = - \frac{1}{L_{\lambda}} (L_{\Theta} \Theta_{\rm v} + L_{\beta} \beta_{\rm v} + L_{\rm v})\\
    \lambda_{0} = - \frac{1}{L_{\lambda}} (L_{\Theta} \Theta_{0} + L_{\beta} \beta_{0} + L_{0})\\
    \mathcal{N} = - (E_{0} + E_{\lambda} \lambda_{0} + E_{\beta} \beta_{0} + E_{\Theta} \Theta_{0})\\
    \mathcal{D} = (E_{\rm v} + E_{\lambda} \lambda_{\rm v} + E_{\beta} \beta_{\rm v} + E_{\Theta} \Theta_{\rm v})
\end{align*}

\section*{Appendix-B: Effect of thermal conduction on the global solutions of inflowing matter}

\begin{figure}[h!]
    \centering
    \includegraphics[width=\columnwidth]{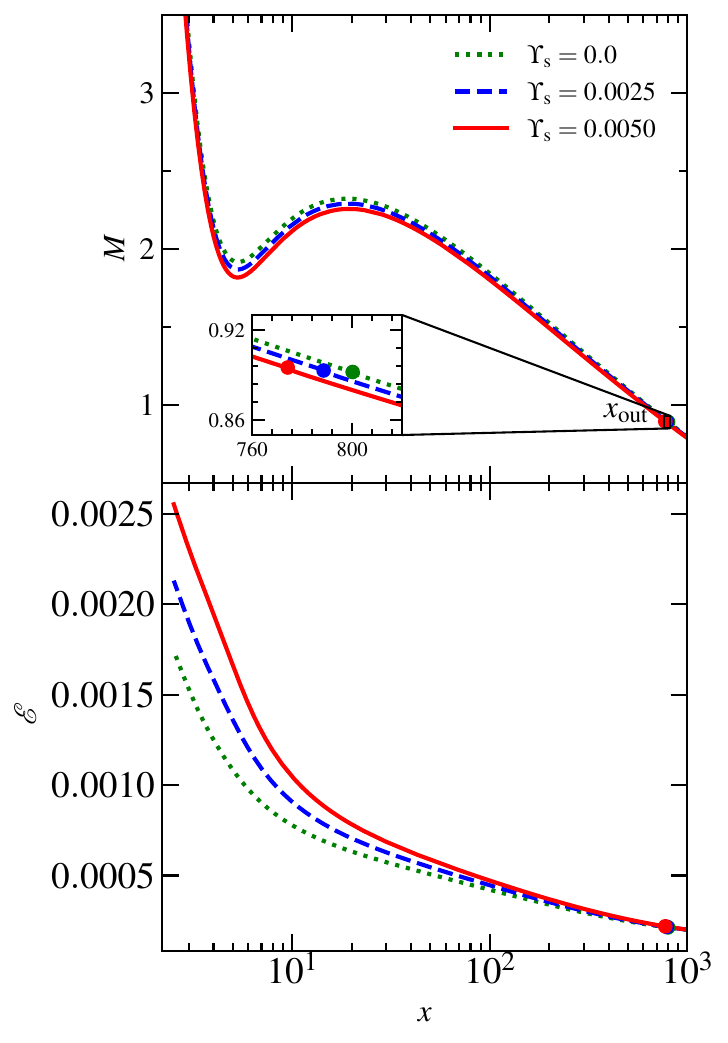}
    \caption{Variation of (a) Mach number ($M= v/C_{\rm s}$) and (b) energy ($\mathcal{E}$) of the inflowing matter with radial distance $(x)$ for $\Upsilon_{\rm s}$  = $0.0$ (green), $0.0025$ (blue) and $0.005$ (red). Here, we choose $a_{\rm k} = 0$, $\alpha_{\rm B} = 0.001$, and fix the input parameters at $x_{\rm edge} = 1000$ as $\lambda_{\rm edge} = 3.20$, $\mathcal{E}_{\rm edge} = 2 \times 10^{-4}$, and $\beta_{\rm edge} = 10^5$. See the text for details.}   
    \label{app1}
\end{figure}

Thermal conduction plays an important role in transferring heat energy within the accretion disk. Hence, it is worthy to examine the effect thermal conduction on the local energy of inflowing matter. In doing so, we choose $a_{\rm k} = 0$, and $\alpha_{\rm B} = 0.001$, and inject matter from the outer edge of the disk at $x_{\rm edge} = 1000$ with $\lambda_{\rm edge} = 3.20$, $\mathcal{E}_{\rm edge} = 2 \times 10^{-4}$, and $\beta_{\rm edge} = 10^5$, respectively. With this, we compute the global accretion solution that passes through the outer critical point at $x_{\rm out} = 800.166$, connecting the outer edge ($x_{\rm edge} = 10^3$) to the event horizon ($x_{\rm h}$) in the absence of thermal conduction ($\Upsilon_{\rm s} = 0$). Thereafter, we calculate the energy ($\mathcal{E}$) profile of the inflow solution. The obtained results are shown in Fig. \ref{app1}, where the variation of (a) Mach number ($M=v/C_{\rm s}$) and (b) local energy ($\mathcal{E}$) are plotted as function of radial coordinate ($x$). In both panels, the dotted (green) curve denotes the results for $\Upsilon_{\rm s} = 0$. Next, we gradually increase $\Upsilon_{\rm s}$ while keeping the other inflow parameters fixed for a non-rotating black hole ($a_{\rm k}=0.0$) and obtain the global inflow solutions. The dashed (blue) and solid (red) curves represent results corresponding to $\Upsilon_{\rm s} = 0.0025$, and $0.005$, respectively. The respective outer critical points are located at $x_{\rm out}=788.559$ and $774.200$, respectively. Evidently, thermal conduction influences the global accretion solutions, leading to an increase in the local energy of the flow.


\end{document}